\documentclass[12pt]{article}
\usepackage{graphicx} 
\usepackage{amsmath}
\usepackage{amssymb}
\usepackage{mathtools}
\usepackage{here} 
\usepackage{hyperref}
\hypersetup{
setpagesize = false,
bookmarksnumbered = true,
bookmarksopen = true,
colorlinks = true,
linkcolor = blue,
citecolor = blue,
}
\usepackage[hang,small,bf]{caption} 
\usepackage[subrefformat=parens]{subcaption}
\usepackage[margin=30truemm]{geometry} 
\numberwithin{equation}{section} 
\usepackage{cite} 

\title{
\begin{flushright}
\ \\*[-80pt]
\begin{minipage}{0.2\linewidth}
\normalsize
HUPD-2501 \\*[40pt] 
\end{minipage}
\end{flushright}
{\Large \bf
Classification of $T^2/Z_m$ orbifold boundary conditions in $SO(N)$ gauge theories
\\*[20pt]}
}

\author{\centerline{
Kota Takeuchi$\,^{1}$\footnote{k-takeuchi@hiroshima-u.ac.jp}$\,\,$
Tomohiro Inagaki$\,^{1,2,3}\,$\footnote{inagaki@hiroshima-u.ac.jp
}}\\*[20pt]
\centerline{
\begin{minipage}{\linewidth}
\begin{center}
$^1${\it \normalsize
Graduate School of Advanced Science and Engineering, Hiroshima University,
Higashi-Hiroshima~739-8526,~Japan \\*[5pt]
$^2${\it \normalsize
Information Media Center, Hiroshima University, Higashi-Hiroshima 739-8521, Japan} \\
$^3${Core of Research for the Energetic Universe, Hiroshima University, Higashi-Hiroshima 739-8526, Japan}
}
\end{center}
\end{minipage}}
\\*[50pt]}
\date{}

\begin{document}
\maketitle

\begin{abstract}
    We generally classify the equivalence classes of the $T^2/Z_m$ $(m=2,3,4,6)$ orbifold boundary conditions (BCs) for the $SO(N)$ gauge group.
    Higher-dimensional gauge theories are defined by gauge groups, matter field contents, and the BCs.
    The numerous patterns of the BCs are classified into a finite number of equivalence classes, each of which consists of the physically equivalent BCs.
    In this paper, we reconstruct the canonical forms of the BCs for the $SO(N)$ gauge group through the ``re-orthogonalization method."
    All the possible equivalent relations between the canonical forms are examined by using the trace conservation laws.
    The number of the equivalence classes in each orbifold model is obtained.
\end{abstract}

\newpage
\tableofcontents
\setcounter{tocdepth}{2}

\section{Introduction} \label{sec_intro}
A higher-dimensional gauge theory compactified on an orbifold is one of the favored theories beyond the Standard Model (SM).
Such an orbifold model realizes the four-dimensional (4D) chiral theories with various patterns of residual gauge symmetries at a low-energy scale\cite{10.1143/PTP.103.613, 10.1143/PTP.105.999, PhysRevD.64.055003, SCRUCCA2003128, PhysRevD.69.055006}.
For instance, gauge-Higgs unification scenario is studied actively\cite{MANTON1979141, FAIRLIE197997, DBFairlie_1979, HOSOTANI1983309, HOSOTANI1983193, HOSOTANI1989233, PhysRevD.109.115005, Maru_Nago_2024}.
In this scenario, the Higgs field is identified with the extra component of the higher-dimensional gauge field.
As the Higgs is protected by the gauge symmetry, the hierarchy problem in SM is solved without supersymmetry\cite{doi:10.1142/S021773239800276X}.
In addition, this model has strong predictive power by integrating the Higgs potential and Yukawa interactions into the gauge terms\cite{doi:10.1142/S0217732302008988}.

In higher-dimensional theories, the boundary conditions (BCs) of fields in the extra-dimensional direction play a key role, significantly affecting the 4D effective theories.
The selection of the BCs determines the patterns of the gauge symmetry breaking and the mass spectra\cite{doi:10.1142/S0217732302008988}.
However, there are numerous choices of the BCs, that lead to the problem of which type of the BCs should be chosen without relying on phenomenological information.
This is called the arbitrariness problem of the BCs\cite{Hosotani:2003ay, HABA2003169, 10.1143/PTP.111.265}.
If our world is located on an effective 4D spacetime embedded in higher-dimensions, there should be a mechanism or principle that selects one BC to describe our world from many choices.
When this arbitrariness problem is resolved, the higher-dimensional theory would become a more convincing unified theory.

The countless choices of the BCs are classified into a finite number of equivalence classes (ECs) through gauge transformations\cite{HABA2003169}. 
Each class consists of the physically equivalent BCs.
The physics depends on the ECs, not on the BCs themselves\cite{HOSOTANI1989233}.%
\footnote{The physical symmetry is determined by the combination of the BCs and the Aharonov-Bohm (AB) phase, based on the Hosotani mechanism\cite{HOSOTANI1983309, HOSOTANI1983193, HOSOTANI1989233}.
The value of the AB phase is almost uniquely fixed by the BCs and the matter field contents.}
Classifying the ECs is not only the important first step in solving the arbitrariness problem of the BCs, but also provides a systematic understanding of the theory, which is useful for model building.

Many works have been done to study the ECs\cite{HABA2003169, 10.1143/PTP.111.265, PhysRevD.69.125014, 10.1143/PTP.120.815, 10.1143/PTP.122.847, doi:10.1142/S0217751X20502061, Kawamura2023,10.1093/ptep/ptae027,10.1093/ptep/ptae082}.
In our previous work, we proposed the trace conservation laws (TCLs) and achieved the general classification of the ECs in the $S^1/Z_2$ and $T^2/Z_m$ $(m=2,3,4,6)$ orbifold models for the gauge group $G=SU(N)$\cite{10.1093/ptep/ptae027,10.1093/ptep/ptae082}.
The TCLs act as strong necessary conditions when classifying the ECs.
In addition, the TCLs are universally valid regardless of the gauge groups and the shapes of the orbifolds.

This paper shows that the ECs in all the 2D orbifold models for $G=SO(N)$ can also be completely classified using the TCLs.
Despite such orbifold models have been studied phenomenologically\cite{PhysRevD.78.096002, PhysRevD.79.079902, PhysRevD.99.095010, PhysRevD.104.115018}, the characterization of the ECs has not been done at all.%
\footnote{Such models are often discussed in the Randall-Sundrum spacetime.
We focus on examining the ECs in flat spacetime since their classification is expected to be independent of the metric.}
In the case of $SO(N)$, the canonical forms of the representation matrices for the BCs need to be cleverly constructed.
This is because the representation matrices for $G=SO(N)$ generally cannot be diagonalized by orthogonal transformations while the ones for $G=SU(N)$ always can be diagonalized by unitary transformations.
We use the ``re-orthogonalization method," which reconstructs the canonical forms for $G=SO(N)$ from the ones for $G=SU(N)$.

This paper is organized as follows.
In Section \ref{sec_general}, we first review the geometric properties of the $T^2/Z_m$ $(m=2,3,4,6)$ orbifolds.
Next the ECs and the TCLs are introduced.
In Section \ref{sec_re-orth}, we explain the re-orthogonalization method and apply to the gauge transformations.
In Section \ref{sec_class}, we classify the ECs in each 2D orbifold model.
Section \ref{sec_concl} is devoted to the conclusion and the discussion.

\section{General properties of \texorpdfstring{$T^2/Z_m$}{T2/Zm} orbifolds} \label{sec_general}
\subsection{Geometric symmetry}
\begin{figure}[t]
    \begin{tabular}{cc}
      \begin{minipage}[t]{0.45\hsize}
        \centering
        \includegraphics[keepaspectratio, scale=0.9]{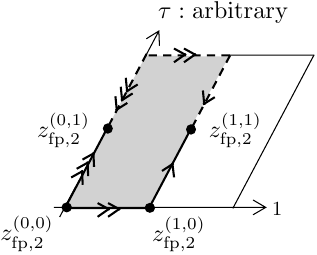}
        \subcaption{$T^2/Z_2$}
        \label{fig_T2Z2}
      \end{minipage} &
      \begin{minipage}[t]{0.45\hsize}
        \centering
        \includegraphics[keepaspectratio, scale=0.9]{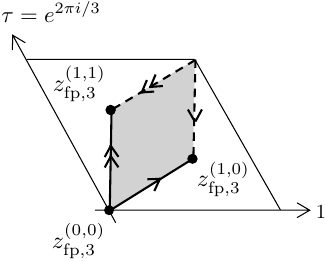}
        \subcaption{$T^2/Z_3$}
        \label{fig_T2Z3}
      \end{minipage} \\ &\\
      \begin{minipage}[t]{0.45\hsize}
        \centering
        \includegraphics[keepaspectratio, scale=0.9]{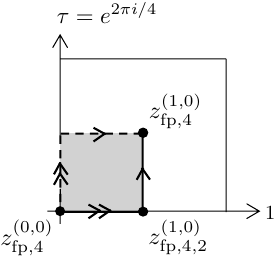}
        \subcaption{$T^2/Z_4$}
        \label{fig_T2Z4}
      \end{minipage} &
      \begin{minipage}[t]{0.45\hsize}
        \centering
        \includegraphics[keepaspectratio, scale=0.9]{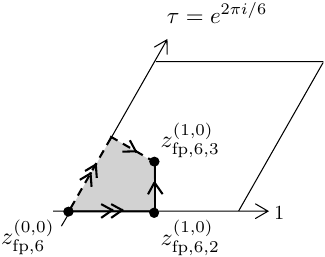}
        \subcaption{$T^2/Z_6$}
        \label{fig_T2Z6}
      \end{minipage} 
    \end{tabular}
     \caption{The shaded areas represent the fundamental regions of $T^2/Z_m$, including the solid lines but excluding the dashed lines. The dots indicate the fixed points.}
     \label{fig_T2Zm}
  \end{figure}

Let $x$ be coordinates of 4D Minkowski spacetime and $z$ be a dimensionless complex coordinate in the 2D extra dimensions, scaled by the length of the extra dimensions.
$T^2/Z_m$ orbifolds are defined by the 2D torus $T^2$ identification and the cyclic group $Z_m$ identification under the following operators:
\begin{equation}
    \hat{\mathcal{T}}_1: z \to z+1, \quad
    \hat{\mathcal{T}}_2: z \to z+\tau, \quad
    \hat{\mathcal{R}}_m: z \to \rho_m z,
\end{equation}
where $\rho_m=e^{2\pi i/m}$, and $\tau\in\mathbb {C}$ $(\text{Im}(\tau)>0,\,|\tau|=1)$ is the complex structure modulus of $T^2$. 
The 2D orbifold is restricted to $m=2,3,4,6$ because of the crystallographic analysis \cite{crystal2020}.
In addition, $\tau$ is limited to $\tau=\rho_m$ for $m=3,4,6$, whereas arbitrary for $m=2$.
It should be noted that $\hat{\mathcal{T}}_2$ is not independent due to $\hat{\mathcal{T}}_2=\hat{\mathcal{R}}_m \hat{\mathcal{T}}_2 \hat{\mathcal{R}}_m^{-1}$ for $m=3,4,6$.
The fundamental regions of $T^2/Z_m$ are given in Fig.\ref{fig_T2Zm}.

\vskip\baselineskip
\vskip0.5\baselineskip

A key feature of orbifolds is the existence of fixed points, which are invariant under the discrete rotations in the compact space.
The $Z_m$-fixed points on $T^2/Z_m$, $z^{(n_1,n_2)}_{\text{fp},m}$, are defined as the unchanged points under the following operators:
\begin{equation} \label{rot_m}
    \hat{\mathcal{T}}_1^{n_1}\hat{\mathcal{T}}_2^{n_2}\hat{\mathcal{R}}_m:
    z \to  \rho_m z + n_1 + n_2 \tau 
    \quad (n_1,n_2\in\mathbb{Z}). 
\end{equation}
These are explicitly written as
\begin{equation}
\begin{alignedat}{2}
    &z_{\text{fp},2}^{(n_1,n_2)} = \frac{n_1 + n_2\tau}{2},
    &&z_{\text{fp},3}^{(n_1,n_2)} = \frac{(2n_1-n_2) + (n_1 + n_2)\tau}{3},\\
    &z_{\text{fp},4}^{(n_1,n_2)} = \frac{(n_1 -n_2) + (n_1 + n_2)\tau}{2},\quad
    &&z_{\text{fp},6}^{(n_1,n_2)} = -n_2+(n_1 + n_2)\tau.
\end{alignedat}
\end{equation}
In addition, the orbifold $T^2/Z_4$ contains the $Z_2$-fixed points $z^{(n_1,n_2)}_{\text{fp},4,2}$, and $T^2/Z_6$ contains the $Z_2$-fixed points $z^{(n_1,n_2)}_{\text{fp},6,2}$ and the $Z_3$-fixed points $z^{(n_1,n_2)}_{\text{fp},6,3}$.
The symmetric operators and the coordinates of them are given as
\begin{alignat}{2}
    &\hat{\mathcal{T}}_1^{n_1}\hat{\mathcal{T}}_2^{n_2}\hat{\mathcal{R}}_4^2:
    z \to  \rho_2 z + n_1 + n_2 \tau,\quad
    &&z_{\text{fp},4,2}^{(n_1,n_2)} = \frac{n_1 + n_2 \tau}{2},\label{rot_4.2}\\
    &\hat{\mathcal{T}}_1^{n_1}\hat{\mathcal{T}}_2^{n_2}\hat{\mathcal{R}}_6^3:
    z \to  \rho_2 z + n_1 + n_2 \tau,\quad
    &&z_{\text{fp},6,2}^{(n_1,n_2)} = \frac{n_1 + n_2 \tau}{2},\label{rot_6.2}\\
    &\hat{\mathcal{T}}_1^{n_1}\hat{\mathcal{T}}_2^{n_2}\hat{\mathcal{R}}_6^2:
    z \to  \rho_3 z + n_1 + n_2 \tau,\quad
    &&z_{\text{fp},6,3}^{(n_1,n_2)} = \frac{(n_1 -n_2) + (n_1 + 2n_2) \tau}{2}\label{rot_6.3}.
\end{alignat}
Note that each fixed point on the covering space is uniquely specified by an integer pair $(n_1,n_2)$.
The symmetric operators (\ref{rot_m}) ((\ref{rot_4.2}), (\ref{rot_6.2}), (\ref{rot_6.3})) represent the $Z_m$  ($Z_p$) rotations around the fixed points $z^{(n_1,n_2)}_{\text{fp},m}$ ($z^{(n_1,n_2)}_{\text{fp},m,p}$).
Let us define $\hat{\mathcal{R}}_m^{(n_1,n_2)}\equiv\hat{\mathcal{T}}_1^{n_1}\hat{\mathcal{T}}_2^{n_2}\hat{\mathcal{R}}_m$ and $\hat{\mathcal{R}}_{m,p}^{(n_1,n_2)}\equiv\hat{\mathcal{T}}_1^{n_1}\hat{\mathcal{T}}_2^{n_2}\hat{\mathcal{R}}^{m/p}_m$ $((m,p)=(4,2),(6,3),(6,2))$.
These satisfy the following consistency conditions:
\begin{equation} \label{consis_rot}
    \hat{\mathcal{R}}_m^{(n_1,n_2)\,m}=(\hat{\mathcal{T}}^{n_1}_1 \hat{\mathcal{T}}^{n_2}_2\hat{\mathcal{R}}_m)^m =1,\quad
    \hat{\mathcal{R}}_{m,p}^{(n_1,n_2)\,p}=(\hat{\mathcal{T}}^{n_1}_1 \hat{\mathcal{T}}^{n_2}_2\hat{\mathcal{R}}^{m/p}_m)^p =1,
\end{equation}
where $1$ denotes the identity operator.

For $m=3,4,6$, $\hat{\mathcal{T}}_2$ is rewritten by $\hat{\mathcal{T}}_2 = \hat{\mathcal{R}}_m \hat{\mathcal{T}}_1 \hat{\mathcal{R}}^{-1}_m$ due to $\tau=\rho_m$.
It is convenient to introduce the additional translation operators:
\begin{equation}
    \hat{\mathcal{T}}_i: z \to z+ \tau^{i-1} \quad\text{for}\,\,i=1,2,\cdots,m.
\end{equation}
These are generated by $\hat{\mathcal{T}}_i \equiv \hat{\mathcal{R}}_m^{i-1} \hat{\mathcal{T}}_{1} \hat{\mathcal{R}}^{1-i}_m$, and commute with each other, i.e., $[\hat{\mathcal{T}}_i,\hat{\mathcal{T}}_j]=0$.
Moreover, the following consistency conditions are satisfied:
\begin{align}
    &\text{for}\,\,m=3:\quad \label{consis_trans3}
    \sum^3_{i=1} \hat{\mathcal{T}}_i =1,\\
    &\text{for}\,\,m=4:\quad \label{consis_trans4}
    \sum^4_{i=1} \hat{\mathcal{T}}_i =1,\quad
    \hat{\mathcal{T}}_i \hat{\mathcal{T}}_{i+2} =1,\\
    &\text{for}\,\,m=6:\quad \label{consis_trans6}
    \sum^6_{i=1} \hat{\mathcal{T}}_i =1,\quad
    \hat{\mathcal{T}}_i \hat{\mathcal{T}}_{i+3} =1,\quad
    \hat{\mathcal{T}}_i \hat{\mathcal{T}}_{i+2} \hat{\mathcal{T}}_{i+4}=1.
\end{align}

Most of the above symmetric operators are not independent.
In this paper, we take $(\hat{\mathcal{R}}_0,\hat{\mathcal{R}}_1,\hat{\mathcal{R}}_2)$ for $m=2$ and $(\hat{\mathcal{R}}_0,\hat{\mathcal{R}}_1)$ for $m=3,4,6$ as the independent bases, redefined by
\begin{align}
    &\text{for}\,\,m=2:\quad
    \hat{\mathcal{R}}_0\equiv \hat{\mathcal{R}}^{(0,0)}_2,\quad
    \hat{\mathcal{R}}_1\equiv \hat{\mathcal{R}}^{(1,0)}_2=\hat{\mathcal{T}}_1 \hat{\mathcal{R}}_0,\quad
    \hat{\mathcal{R}}_2\equiv \hat{\mathcal{R}}^{(0,1)}_2=\hat{\mathcal{T}}_2 \hat{\mathcal{R}}_0,\\
    &\text{for}\,\,m=3:\quad
    \hat{\mathcal{R}}_0\equiv \hat{\mathcal{R}}^{(0,0)}_3,\quad
    \hat{\mathcal{R}}_1\equiv \hat{\mathcal{R}}^{(1,0)}_3=\hat{\mathcal{T}}_1 \hat{\mathcal{R}}_0,\\
    &\text{for}\,\,m=4:\quad
    \hat{\mathcal{R}}_0\equiv \hat{\mathcal{R}}^{(0,0)}_4,\quad
    \hat{\mathcal{R}}_1\equiv \hat{\mathcal{R}}^{(1,0)}_4=\hat{\mathcal{T}}_1 \hat{\mathcal{R}}_0,\\
    &\text{for}\,\,m=6:\quad
    \hat{\mathcal{R}}_0\equiv \hat{\mathcal{R}}^{(0,0)}_6,\quad
    \hat{\mathcal{R}}_1\equiv \hat{\mathcal{R}}^{(1,0)}_{6,3}=\hat{\mathcal{T}}_1 \hat{\mathcal{R}}_0^2.
\end{align}
Note that only two bases are required for $m=3,4,6$ since $\hat{\mathcal{T}}_2=\hat{\mathcal{R}}_0 \hat{\mathcal{T}}_1 \hat{\mathcal{R}}_0^{-1}$.
Their consistency conditions are summarized in Table \ref{tab_consis}.
The basic conditions in Table \ref{tab_consis} lead to all the consistency conditions (\ref{consis_rot}), (\ref{consis_trans3}), (\ref{consis_trans4}), (\ref{consis_trans6}) and $[\hat{\mathcal{T}}_i,\hat{\mathcal{T}}_j]=0$ \cite{10.1093/ptep/ptae082}.
The reason for choosing the rotation operators as the bases, rather than the translation operators, is that they possess several invariant quantities under ``BCs-connecting gauge transformations," as will be explained next.

\begin{table}[t]
    \centering
    \scalebox{0.95}{
    \renewcommand{\arraystretch}{1.8}
    \setlength{\tabcolsep}{10pt}
    \begin{tabular}{cccccc} 
    \hline \hline
    $T^2/Z_m$ & The bases & \multicolumn{4}{c}{The basic consistency conditions} \\
    \hline
    $T^2/Z_2$ 
    &$(\hat{\mathcal{R}}_0, \hat{\mathcal{R}}_1, \hat{\mathcal{R}}_2)$
    &$\hat{\mathcal{R}}_0^2=1,$
    &$\hat{\mathcal{R}}_1^2=1,$
    &$\hat{\mathcal{R}}_2^2=1,$
    &$(\hat{\mathcal{R}}_1 \hat{\mathcal{R}}_0 \hat{\mathcal{R}}_2)^2=1$ \\
    $T^2/Z_3$ 
    &$(\hat{\mathcal{R}}_0, \hat{\mathcal{R}}_1)$
    &$\hat{\mathcal{R}}_0^3=1,$
    &$\hat{\mathcal{R}}_1^3=1,$
    &$(\hat{\mathcal{R}}_1 \hat{\mathcal{R}}_0)^3=1$&\\
    $T^2/Z_4$ 
    &$(\hat{\mathcal{R}}_0, \hat{\mathcal{R}}_1)$
    &$\hat{\mathcal{R}}_0^4=1,$
    &$\hat{\mathcal{R}}_1^4=1,$
    &$(\hat{\mathcal{R}}_1 \hat{\mathcal{R}}_0)^2=1$&\\
    $T^2/Z_6$ 
    &$(\hat{\mathcal{R}}_0, \hat{\mathcal{R}}_1)$
    &$\hat{\mathcal{R}}_0^6=1,$
    &$\hat{\mathcal{R}}_1^3=1,$
    &$(\hat{\mathcal{R}}_1 \hat{\mathcal{R}}_0)^2=1$&\\
    \hline \hline
    \end{tabular}}
    \caption{The basic consistency conditions}
    \label{tab_consis}
\end{table}

\subsection{Boundary conditions and Equivalence classes} 
Fields on compact extra dimensions follow boundary conditions (BCs) corresponding to the geometric symmetries of the extra space.
Let the bulk field $\Phi(x^\mu,z,\bar{z})$ be a multiplet of a gauge group $G$ on an orbifold, and its symbol $\hat{\mathcal{O}}$ be the symmetric operators, such as $\hat{\mathcal{R}}^{(n_1,n_2)}_m$, $\hat{\mathcal{R}}^{(n_1,n_2)}_{m,p}$ and $\hat{\mathcal{T}}_i$.
Then, its BCs are generically written as
\begin{equation}
    \Phi(x^\mu, \hat{\mathcal{O}}(z), \hat{\mathcal{O}}(\bar{z})) = T_\Phi (\hat{\mathcal{O}})\, \Phi(x^\mu,z,\bar{z}),
\end{equation}
where $T_\Phi (\hat{\mathcal{O}})$ represents an appropriate representation matrix, determined by the $\hat{\mathcal{O}}$ invariance of its Lagrangian.
In the $T^2/Z_m$ orbifold models, $T_\Phi (\hat{\mathcal{R}}_i)$ $(i=0,1,2)$ are almost uniquely characterized by $R_i$, which are representation matrices of $G$.%
\footnote{$T_\Phi (\hat{\mathcal{R}}_i)$ are determined by $R_i$ and intrinsic phases factor\cite{doi:10.1142/S0217751X20502061}.}
They satisfy the consistency conditions by replacing $\hat{\mathcal{R}}_i$ with $R_i$ in Table \ref{tab_consis}.
There are numerous choices of $R_i$, indicating the arbitrariness of the BCs.
The different choices of the BCs lead to the different low-energy effective 4D theories\cite{doi:10.1142/S0217732302008988}.

Is there a way to systematize the selection of BCs? 
Using gauge transformations, the numerous patterns of the BCs are classified into a finite number of equivalence classes (ECs), each of which consists of the physically equivalent BCs\cite{HABA2003169}.
Let $\hat{\mathcal{R}}$ be a $\rho$ rotation operator around a fixed point $z_{\text{fp}}$.
The BC of the field $\Phi$ with respect to $\hat{\mathcal{R}}$ is described by
\begin{equation} \label{BC_R}
    \Phi(x^\mu, \hat{\mathcal{R}}(z), \hat{\mathcal{R}}(\bar{z})) = T_\Phi (\hat{\mathcal{R}})\, \Phi(x^\mu,z,\bar{z}),
\end{equation}
with
\begin{equation}
    \hat{\mathcal{R}}(z): z\to \rho(z-z_{\text{fp}})+z_{\text{fp}}.
\end{equation}
When $\Phi$ is gauge transformed to $\Phi'$, the BC of $\Phi'$ is written as
\begin{equation}
    \Phi'(x^\mu, \hat{\mathcal{R}}(z), \hat{\mathcal{R}}(\bar{z})) = T_{\Phi'} (\hat{\mathcal{R}})\, \Phi'(x^\mu,z,\bar{z}).
\end{equation}
Define $R$ and $R'$ as representation matrces of $G$, which characterize $T_{\Phi} (\hat{\mathcal{R}})$ and $T_{\Phi'} (\hat{\mathcal{R}})$.
They obey the following relation:
\begin{equation} \label{GT_R}
    R' = \Omega(x^\mu, \hat{\mathcal{R}}(z)) R\, \Omega^{-1}(x^\mu, z),
\end{equation}
where $\Omega(x^\mu,z)$ is a transformation matrix.
$R'$ is generally coordinate-dependent, but if $R'$ remains constant and satisfies the same consistency conditions as $R$, then $R'$ can be regarded as another choice of $R$.
We will refer to Eq.(\ref{GT_R}) as the ``BCs-connecting gauge transformation" in this paper.
When $R$ and $R'$ are connected by such a transformation, they yield equivalent physics:
\begin{equation}
    R \sim R'.
\end{equation}

\begin{figure}[t]
\centering\scalebox{0.85}{\includegraphics*{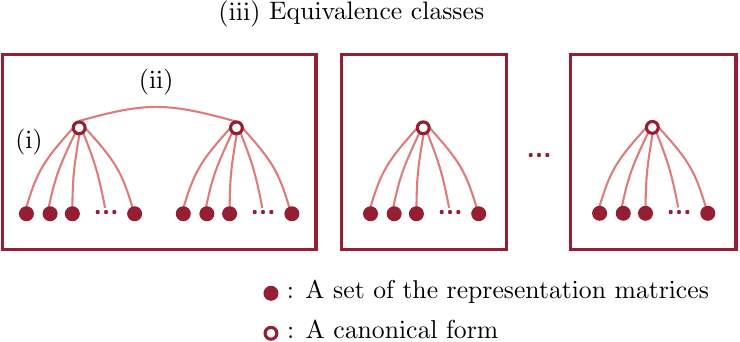}}
\caption{A schematic diagram of ECs is shown.
The filled and open points denote the sets of the representation matrices for the BCs and their canonical forms, respectively.
The lines represent gauge transformations and the square boxes indicate the ECs.}
\label{fig_ECs}
\end{figure}

Classifying the ECs plays a crucial role in solving the arbitrariness problem of the BCs and is valuable for model building because we obtain a systematic understanding of the theory.
As illustrated in Fig.\ref{fig_ECs}, the following three questions should be considered:
\begin{itemize}
    \item[(i)\,\,] 
    What are the canonical forms of the representation matrices for BCs?
    \item[(ii)\,] 
    Which types of the canonical forms are connected by gauge transformations?
    \item[(iii)] 
    How many ECs exist?
\end{itemize}
We will answer these questions for each orbifold model in Section \ref{sec_class}.

\subsection{Trace Conservation Laws}
To classify the ECs generally, gauge invariants obtained from the trace conservation laws (TCLs) are needed.
Let us consider $R$'s gauge transformation (\ref{GT_R}).
Typically, the traces of $R$ and $R'$ are not the same because $\Omega$ and $\Omega^{-1}$ on the right-hand side of Eq.(\ref{GT_R}) have the different arguments.
Nevertheless, their traces are always equal under the BCs-connecting gauge transformations.
The reason is as follows.
$R'$ remains constant under the BCs-connecting gauge transformations, so that the entire right-hand side of Eq.(\ref{GT_R}) is also coordinate-independent.
It means that if the traces of $R$ and $R'$ coincide at a particular point, they must equal over the whole complex plane.
In fact, the trace is conserved at the fixed point $z_{\text{fp}}$:
\begin{equation}
    \begin{aligned}
    \mathrm{tr} R' |_{z=z_{\text{fp}}}
    &= \mathrm{tr} \left[ \Omega(\hat{\mathcal{R}}(z_{\text{fp}})) R\, \Omega^{-1}(z_{\text{fp}}) \right]\\
    &= \mathrm{tr} \left[ \Omega(z_{\text{fp}}) R\, \Omega^{-1}(z_{\text{fp}}) \right] \\
    &= \mathrm{tr} \left[\Omega^{-1}(z_{\text{fp}})\, \Omega(z_{\text{fp}}) R \right] \\
    &= \mathrm{tr} R,
    \end{aligned}
\end{equation}
where we use $\mathrm{tr}(ABC)=\mathrm{tr}(CAB)$.
(Hereafter $\Omega$'s argument $x^\mu$ will be omitted.)
As a result, it is concluded that the traces of $R$ and $R'$ are always equal under the BCs-connecting gauge transformations, which is called the ``Trace conservation law (TCL)."%
\footnote{Similarly, the determinant of $R$ is conserved, but the TCLs are more powerful necessary conditions for classifying BCs.
If the trace is conserved, the determinant is also conserved in many cases.}
In other words, such traces are gauge invariant quantities for the BCs-connecting gauge transformations.

The TCLs act as strong necessary conditions, significantly narrowing down the possible equivalent relations between BCs.
They enable the BCs to be classified without explicitly specifying the forms of gauge transformations.
In addition, the TCLs are valid for any gauge group and orbifold type.
In our previous papers\cite{10.1093/ptep/ptae027, 10.1093/ptep/ptae082}, we completed the general classification of the BCs in the $T^2/Z_m$ orbifold models for the $SU(N)$ gauge group using the TCLs.
In these models, all the traces of the rotation matrices for $\hat{\mathcal{R}}^{(n_1,n_2)}_m$ and $\hat{\mathcal{R}}^{(n_1,n_2)}_{m,p}$ are conserved.
There are numerous gauge invariant quantities corresponding to all the fixed points specified by $(n_1,n_2)$, but if the bases $R_i$ $(i=0,1,2)$ commute with each other, only a few invariant quantities remain independent, given in Table \ref{tab_tr}\cite{10.1093/ptep/ptae082}.

\begin{table}[t]
    \centering
    \scalebox{1.0}{
    \renewcommand{\arraystretch}{1.8}
    \setlength{\tabcolsep}{10pt}
    \begin{tabular}{cccccc}
    \hline \hline
    $T^2/Z_m$ & The bases & \multicolumn{4}{c}{The gauge invariant quantities} \\
    \hline
    $T^2/Z_2$ 
    &$(\hat{\mathcal{R}}_0, \hat{\mathcal{R}}_1, \hat{\mathcal{R}}_2)$
    &$\mathrm{tr} R_0,$
    &$\mathrm{tr} R_1,$
    &$\mathrm{tr} R_2,$
    &$\mathrm{tr} (R_1 R_0 R_2)$ \\
    $T^2/Z_3$ 
    &$(\hat{\mathcal{R}}_0, \hat{\mathcal{R}}_1)$
    &$\mathrm{tr} R_0,$
    &$\mathrm{tr} R_1,$
    &$\mathrm{tr} (R^2_1 R^2_0)$ &\\
    $T^2/Z_4$ 
    &$(\hat{\mathcal{R}}_0, \hat{\mathcal{R}}_1)$
    &$\mathrm{tr} R_0,$
    &$\mathrm{tr} R_1,$
    &$\mathrm{tr} (R_1 R_0),$
    &$\mathrm{tr} R_0^2$ \\
    $T^2/Z_6$ 
    &$(\hat{\mathcal{R}}_0, \hat{\mathcal{R}}_1)$
    &$\mathrm{tr} R_0,$
    &$\mathrm{tr} R_1,$
    &$\mathrm{tr} (R_1 R_0)$ &\\
    \hline \hline
    \end{tabular}}
    \caption{The gauge invariant quantities for the commutative bases}
    \label{tab_tr}
\end{table}

\section{Re-orthogonalization method} \label{sec_re-orth}
In this section, we introduce a convenient method for constructing the canonical forms for the $SO(N)$ group, which is called ``re-orthogonalization method" in this paper. 
First we review this method\cite{re-orth}, and next show that it can be applied to the BCs-connecting gauge transformations.

\subsection{Orthogonal transformations}
An orthogonal matrix cannot be diagonalized by an orthogonal transformation if it has complex eigenvalues.
It can be diagonalized by a unitary transformation, but the resulting diagonal matrix generally loses orthogonality.
How far is an orthogonal matrix be reduced while keeping its orthogonality?

Let $R$ be an $O(N)$ matrix with $n_\pm$ real eigenvalues $\pm1$ and $2m$ complex eigenvalues (phase factors).
$R$ is diagonalized to a diagonal matrix $R_\text{d}$ by a unitary matrix $U$:
\begin{equation}
    R= U R_\text{d} U^\dagger, \quad 
    R_\text{d}=\mathrm{diag}\, 
    (\underbrace{1, \cdots, 1}_{n_+}, 
    \underbrace{-1, \cdots, -1}_{n_-}, 
    \underbrace{\alpha_1, \bar{\alpha}_1, \cdots, \alpha_m, \bar{\alpha}_m}_{2m}).
\end{equation}
Note that when $R$ has a complex eigenvalue, it must also have its conjugate eigenvalue because $R$'s characteristic polynomial is real.
The unitary matrix $U$ can be written as
\begin{equation} \label{eigV}
    U =  
    (\mathbf{t}^+_1, \cdots, \mathbf{t}^+_{n_+},
    \mathbf{t}^-_1, \cdots, \mathbf{t}^-_{n_-},
    \mathbf{u}_1, \bar{\mathbf{u}}_1,\cdots,\mathbf{u}_m, \bar{\mathbf{u}}_m),
\end{equation}
where $\mathbf{t}^{\pm}_i$ and $\mathbf{u}_j$ are real and complex orthonormal bases and $\bar{\mathbf{u}}_j$ is the conjugate vector of $\mathbf{u}_j$.%
\footnote{$R$ can be diagonalized to $R_\text{d}$ by another unitary matrix $U'$, but $U'$ can always be converted to $U$ by a suitable basis transformation $W$ while keeping $R_\text{d}$ unchanged: $R=UR_\text{d} U^\dagger=(UW) (W^\dagger R_\text{d} W) (W^\dagger U^\dagger)=U'R_\text{d} U'^\dagger$.}
Let $\mathbf{u}_j$ be written as $\mathbf{u}_j= (\mathbf{a}_j - i \mathbf{b}_j)/\sqrt{2}$, where $\mathbf{a}_j$ and $\mathbf{b}_j$ are real orthonormal vectors.
The bases $(\mathbf{u}_j,\bar{\mathbf{u}}_j)$ are transformed into $(\mathbf{a}_j,\mathbf{b}_j)$ by the following $2\times 2$ unitary matrix $V$:
\begin{equation} \label{Re_V}
    (\mathbf{u}_j,\bar{\mathbf{u}}_j) V = (\mathbf{a}_j,\mathbf{b}_j),\quad
    V=\frac{1}{\sqrt{2}}\begin{pmatrix} 1&i\\1&-i \end{pmatrix}.
\end{equation}
In this basis, the diagonal matrix is re-orthogonalized to a rotation matrix:
\begin{equation}
    V^\dagger \begin{pmatrix} \alpha_j &\\& \bar{\alpha}_j \end{pmatrix} V
    = \begin{pmatrix} \cos{\theta_j}&-\sin{\theta_j}\\ \sin{\theta_j}&\cos{\theta_j} \end{pmatrix}\equiv r(\theta_j) \in {O(2)},
\end{equation}
where $\alpha_j=\cos{\theta_j}+i\sin{\theta_j}$ $(\theta_j \in\mathbb{R})$.
As a result, we find that $R$ is block-diagonalized to the following canonical form $R_{\text{bd}}$:
\begin{equation}
    R=U R_\text{d} U^\dagger= O R_\text{bd} O^\top,\quad 
    R_\text{bd}=I_{n_+} \oplus (-I_{n_-}) \oplus 
    r(\theta_1)\oplus\cdots\oplus r(\theta_m),
\end{equation}
where $I_n$ is the $n\times n$ unit matrix and $O$ is an orthogonal matrix defined by
\begin{align}
    O&\equiv U\, (I_{n_+}\oplus I_{n_-}\oplus \underbrace{V\oplus\cdots\oplus V}_{m})\\
    &=(\mathbf{t}_1, \cdots, \mathbf{t}_{n_+},
    \mathbf{t}_1, \cdots, \mathbf{t}_{n_-},
    \mathbf{a}_1,\mathbf{b}_1,\cdots,\mathbf{a}_m,\mathbf{b}_m).
\end{align}
It is easily checked that the new bases $\{\mathbf{t}_i,\mathbf{a}_j,\mathbf{b}_j\}$ are orthonormal.

\subsection{Gauge transformations}
We show that the re-orthogonalization method can be applied to the BCs-connecting gauge transformations which simultaneously diagonalize the representation matrices in the orbifold models. 
According to Ref.\!\cite{Kawamura2023}, in the $T^2/Z_m$ models for $G=SU(N)$, any representation matrices $R_i$ are first reduced to the direct-sum of the following $m\times m$ blocks $r_i$ by (global) unitary transformations $U$:
\begin{equation} \label{r_blocks}
    r_0=X,\quad 
    r_1= \left(\sum^m_{i=1} a_i Y^i \right) X,
\end{equation}
where
\begin{equation}
    X=
    \begin{pmatrix}
        \rho_m &&&\\
        &\rho_m^2 &&\\
        &&\ddots &\\
        &&&1
    \end{pmatrix},\quad
    Y=
    \begin{pmatrix}
        &&&1 \\
        1 &&&\\
        &\ddots&& \\
        &&1&
    \end{pmatrix},
\end{equation}
and $a_i$ are real or complex numbers.
For $m=2$, $Y$ is equal to the Pauli matrices $\sigma_1$, and $r_2$ is given the same form as $r_1$ except for the parameters $a_i$.
These blocks can be simultaneously diagonalized by the gauge transformations with
\begin{equation} \label{omega}
    \Omega(z) = \exp{\left[ i(\beta zY +\bar{\beta}\bar{z}Y^\dagger)\right]}, 
\end{equation}
where $\beta$ is an appropriate complex number.%
\footnote{For $m=4,6$, some blocks remain off-diagonal, which will be discussed in Section \ref{sec_class}.}

The re-orthogonalization method can be applied to the gauge transformations with Eq.(\ref{omega}).
This is because they are equivalent to the following unitary transformations:
\begin{equation} \label{gauge=uni}
\begin{aligned}
    R'_i &= \Omega(z^{(i)}_{\text{fp}}+\rho z')\, (U R_i U^\dag)\, \Omega^{\dagger} (z^{(i)}_{\text{fp}}+z') \\
    &= \Omega(z^{(i)}_{\text{fp}})\, \Omega(\rho z')\, (U R_i U^\dag)\, \Omega^\dagger(z')\, \Omega^\dagger(z^{(i)}_{\text{fp}}) \\
    &= \Omega(z^{(i)}_{\text{fp}})\, \Omega(\rho z')\, \Omega^\dagger(\rho z')\, (U R_i U^\dag)\, \Omega^\dagger(z^{(i)}_{\text{fp}})\\
    &= \Omega(z^{(i)}_{\text{fp}}) U R_i U^\dag  \Omega^\dagger(z^{(i)}_{\text{fp}}),
\end{aligned}
\end{equation}
where $z=z^{(i)}_{\text{fp}}+z'$ and $z^{(i)}_{\text{fp}}$ denote the $\hat{\mathcal{R}}_i$'s fixed points, and $(U R_i U^\dag)$ are the direct-sum of Eq.(\ref{r_blocks}).
At the second and third equal signs, $\Omega(\alpha+\beta)=\Omega(\alpha) \Omega(\beta)$ and $XY=\rho_m YX$ are used.
Eq.(\ref{gauge=uni}) indicates that $R_i$ for $G=SU(N)$ are diagonalized only by the global unitary transformations $\Omega(z^{(i)}_{\text{fp}}) U$.
Therefore, if all the complex conjugate eigenvalue pairs occupy the same positions in both $R_0$ and $R_1$, i.e., $\overline{(R_0)}_{ii}=(R_0)_{jj} \Leftrightarrow \overline{(R_1)}_{ii}=(R_1)_{jj}$ for all $i,j$, then $R_0$ and $R_1$ are simultaneously re-orthogonalized as
\begin{equation}
\begin{aligned}
    V^\dag R'_i V 
    &= V^\dag \Omega (z^{(i)}_{\text{fp}}+\rho z')\, U R_i U^\dag\, \Omega^\dag (z^{(i)}_{\text{fp}}+z') V \\
    &= (V^\dag \Omega (z^{(i)}_{\text{fp}}+\rho z') V)\, (V^\dag U)\, R_i \,(U^\dag V)\, (V^\dag \Omega^\dag (z^{(i)}_{\text{fp}}+z') V),
\end{aligned}
\end{equation}
where $(V^\dag R'_i V)\in O(N)$ are the block-diagonal orthogonal matrices and $V$ is the direct-sum representation of the $2\times 2$ matrix $V$ in Eq.(\ref{Re_V}).

Finally, it is necessary to confirm that $\tilde{\Omega}(z)\equiv V^\dag \Omega(z) V$ is an element of the $SO(N)$ group.%
\footnote{If $\tilde{\Omega}(z)$ is an orthogonal matrix, $\tilde{U}\equiv V^\dag U$ is also an orthogonal one, which block-diagonalizes $R_0$.}
To show this, we only need to prove that if $\tilde{\Omega}(z)$ is a real at a point $z=z_0$, then it is real throughout $z$.
The transformation matrix $\Omega(z)$ in Eq.(\ref{omega}) is calculated as
\begin{equation}
\begin{aligned}
    \Omega(z) 
    &= W^\dag \exp{\left[ i(\beta zX +\bar{\beta}\bar{z}X^\dag)\right]} W \\
    &= W^\dag \text{diag} \left( e^{2i \text{Re}(\beta z\rho_m)}, e^{2i \text{Re}(\beta z\rho^2_m)}, \cdots, e^{2i \text{Re}(\beta z)} \right) W,
\end{aligned}
\end{equation}
where $W$ is the unitary matrix that diagonalizes $Y$, i.e., $WYW^\dag =X$. 
It is noticed that each element of $\Omega(z)$ is a complex function with real variables:
\begin{equation}
    \Omega(z)_{ij} = \sum_{k=1}^m 
    \alpha_{k,ij} \cos{(2\text{Re}(\beta z\rho^k_m))} +
    \alpha'_{k,ij} \sin{(2\text{Re}(\beta z\rho^k_m))},\quad
    \alpha_{k,ij}, \alpha'_{k,ij} \in \mathbb{C}.
\end{equation} 
If $\tilde{\Omega}(z)\equiv V^\dag \Omega(z) V$ is a real at a point, $z=z_0$, the complex parameters in $\Omega(z)_{ij}$ are replaced by real parameters:
\begin{equation}
    \tilde{\Omega}(z)_{ij} = \sum_{k=1}^m 
    a_{k,ij} \cos{(2\text{Re}(\beta z\rho^k_m))} +
    a'_{k,ij} \sin{(2\text{Re}(\beta z\rho^{k}_m))},\quad
    a_{k,ij}, a'_{k,ij} \in \mathbb{R}.
\end{equation} 
Then $\tilde{\Omega}(z)$ is real throughout $z$, that is, $\tilde{\Omega}(z)$ is an element of the $SO(N)$ group.%
\footnote{$\tilde{\Omega}(z)$ satisfies $|\tilde{\Omega}(z)|=1$ because of $|\Omega(z)|=1$.}

From the above discussion, it is concluded that the canonical forms for $G=SO(N)$ are reconstructed from the diagonal forms for $G=SU(N)$ through the re-diagonalization method if the complex conjugate eigenvalue pairs occupy the same positions in both $R_0$ and $R_1$.
In Section \ref{sec_class}, we will concretely derive the canonical forms in each orbifold model.

\section{Classification of boundary conditions} \label{sec_class}
In this section, we classify the BCs for $G=SO(N)$ in the $T^2/Z_m$ $(m=2,3,4,6)$ orbifold models.
Their canonical forms, the equivalent relations between them and the number of the ECs are derived.

\subsection{\texorpdfstring{$T^2/Z_2$}{T2/Z2} orbifold}
In the $T^2/Z_2$ model, the representation matrices $R_i \in O(N)$ $(i=0,1,2)$ satisfy the following consistency conditions (see Table \ref{tab_consis}):
\begin{equation}
    R_0^2=I_N,\quad R_1^2=I_N,\quad 
    R_2^2=I_N,\quad (R_1 R_0 R_2)^2=I_N.
\end{equation}
Since $R_i$ only have the real eigenvalues $\pm1$, their forms remain unchanged after the re-orthogonalization.
It indicates that the classification results of the BCs for $G=SO(N)$ and $SU(N)$ are exactly the same.%
\footnote{The classification results for $G=SO(N)$ and $SU(N)$ in the $S^1/Z_2$ model are also the same.}
In the case of $SU(N)$, the ECs has already been classified in Ref.\!\cite{10.1093/ptep/ptae082}.
The canonical forms of $(R_0,R_1,R_2)$, their equivalent relations, and the number of the ECs, $S_N$, are summarized in Table \ref{tab_Z6}, where $p,q,\cdots,w$ denote the numbers of each eigenvalue ($p+q+\cdots +w=N$).
\begin{table}[t]
    \centering
    \scalebox{0.9}{
    \renewcommand{\arraystretch}{1.8}
    \setlength{\tabcolsep}{10pt}
    \begin{tabular}{cc} 
    \hline \hline \\[-12pt]
    The canonical forms: &
    $\begin{alignedat}{50}
    R_0&= 
    +I_p &&\oplus +I_q &&\oplus +I_r &&\oplus +I_s &&\oplus
    -I_t &&\oplus -I_u &&\oplus -I_v &&\oplus -I_w \\
    R_1&= 
    +I_p &&\oplus +I_q &&\oplus -I_r &&\oplus -I_s &&\oplus
    +I_t &&\oplus +I_u &&\oplus -I_v &&\oplus -I_w \\
    R_2&= 
    +I_p &&\oplus -I_q &&\oplus +I_r &&\oplus -I_s &&\oplus
    +I_t &&\oplus -I_u &&\oplus +I_v &&\oplus -I_w \\
    \end{alignedat}$ \\[24pt]
    The equivalent relations: &
    $\begin{aligned}
        [\,p,q,r,s \,|\, t,u,v,w \,] 
        &\sim [\,p-1,q+1,r,s \,|\, t,u,v+1,w-1 \,] \\
        &\sim [\,p-1,q,r+1,s \,|\, t,u+1,v,w-1 \,] \\
        &\sim [\,p-1,q,r,s+1 \,|\, t+1,u,v,w-1 \,]
    \end{aligned}$ \\[18pt]
    The total number of the ECs: &
    $S_N= \frac{1}{3} (N+1)^2 (N^2+2N +3)$ \\[12pt]
    \hline \hline
    \end{tabular}}
    \caption{The classification results of the BCs in the $T^2/Z_2$ model with $G=SO(N)$.}
    \label{tab_Z2}
\end{table}

\subsection{\texorpdfstring{$T^2/Z_3$}{T2/Z3} orbifold}
Let us derive the canonical forms of the representation matrices $(R_0,R_1)$ for $G=SO(N)$ in the $T^2/Z_3$ model using the re-orthogonalization method.
$R_i \in O(N)$ $(i=0,1)$ satisfy the following consistency conditions (see Table \ref{tab_consis}):
\begin{equation}
    R_0^3=I_N,\quad R_1^3=I_N,\quad (R_1 R_0)^3=I_N.
\end{equation}
$(R_0, R_1)$ is diagonalized to the following form through unitary and gauge transformations \cite{Kawamura2023}:
\begin{equation} \label{Z3_diag}
\begin{alignedat}{12}
    R_0
    &=\mathrm{diag}\,(&&\omega,\cdots,\, 
        &&\omega,\cdots,\,
        &&\omega,\cdots
        &&|\,&&\bar{\omega},\cdots,\,
        &&\bar{\omega},\cdots,\,
        &&\bar{\omega},\cdots
        &&|\,&&1,\cdots,\, 
        &&1,\cdots,\,
        &&1,\cdots), \\
    R_1
    &=\mathrm{diag}\,(&&\underbrace{\omega,\cdots}_{s_1},\, 
        &&\underbrace{\bar{\omega},\cdots}_{t_1},\,
        &&\underbrace{1,\cdots}_{r_1}
        &&|\,&&\underbrace{\omega,\cdots}_{t_2},\,
        &&\underbrace{\bar{\omega},\cdots}_{s_2},\,
        &&\underbrace{1,\cdots}_{r_2}
        &&|\,&&\underbrace{\omega,\cdots}_{q_1},\, 
        &&\underbrace{\bar{\omega},\cdots}_{q_2},\,
        &&\underbrace{1,\cdots}_{p}),
\end{alignedat}
\end{equation}
where $\omega=e^{2\pi i/3}$ and $p,q_i,r_i,s_i$ $(i=1,2)$ denote the numbers of each eigenvalue.
An orthogonal matrix has the same numbers of conjugate complex eigenvalues.
$R_0$, $R_1$, $(R_0 R_1)$, $(R_0^2 R_1)$, $(R_0 R_1^2)$ and $(R_0^2 R_1^2)$ have the equal numbers of $\omega$ and $\bar{\omega}$, which lead to $q_1=q_2$, $r_1=r_2$, $s_1=s_2$ and $t_1=t_2$.
Since the complex conjugate eigenvalue pairs occupy the same positions in both $R_0$ and $R_1$, the re-orthogonalization method can be applied to Eq.(\ref{Z3_diag}). 
Using the unitary matrix $V$ in Eq.(\ref{Re_V}), the diagonal blocks of $(R_0,R_1)$ in Eq.(\ref{Z3_diag}) are re-orthogonalized to
\begin{equation}
    V^\dag \begin{pmatrix} \omega &0\\0& \bar{\omega} \end{pmatrix} 
    V = r_3 \in O(2), \quad
    V^\dag \begin{pmatrix} \bar{\omega} &0\\0& \omega \end{pmatrix} 
    V = r_{-3} \in O(2),
\end{equation}
where
\begin{equation} \label{r_m}
    r_m \equiv
    \begin{pmatrix}
    \cos{(2\pi/m)} & -\sin{(2\pi/m)} \\
    \sin{(2\pi/m)} & \cos{(2\pi/m)}
    \end{pmatrix}.
\end{equation}
As a result, we obtain the following canonical form of $(R_0,R_1)$ for $G=SO(N)$:
\begin{equation} \label{Z3_StandForm}
\begin{alignedat}{50}
    R_0&= I_p 
    &&\oplus(I_2 \oplus\cdots)
    &&\oplus(r_3 \oplus\cdots)
    &&\oplus(r_3 \oplus\cdots)
    &&\oplus(r_3 \oplus\cdots), \\
    R_1&= I_p 
    &&\oplus \underbrace{(r_3 \oplus\cdots)}_{q}
    &&\oplus \underbrace{(I_2 \oplus\cdots)}_{r}
    &&\oplus \underbrace{(r_3 \oplus\cdots)}_{s}
    &&\oplus \underbrace{(r_{-3} \oplus\cdots)}_{t},
\end{alignedat}
\end{equation}
where $p(=p_i)$, $q(=q_i)$, $r(=r_i)$, $s(=s_i)$ and $t(=t_i)$ denote the numbers of each block.

Next, examine the equivalent relations between the canonical forms through the TCLs.
Since $R_0$ and $R_1$ in the canonical form (\ref{Z3_StandForm}) commute with each other, only the following three traces are independent (see Table \ref{tab_tr}):
\begin{align}
    \mathrm{tr} R_0 &= p+2q-r-s-t, \\
    \mathrm{tr} R_1 &= p-q+2r-s-t,\\
    \mathrm{tr} (R_1 R_0) &=p-q-r-s+2t.
\end{align}
Note that $\mathrm{tr} (R_1^2 R_0^2)$ is equal to $\mathrm{tr} (R_1 R_0)$ for $R_i \in O(N)$.
Given that $p+2q+2r+2s+2t$ is invariant by definition, it follows that $p+2q$, $p+2r$, $p+2t$ and $p-s$ are invariant.
Therefore, it is concluded that the possible transformations are restricted to the following:
\begin{equation} \label{Z3_equiv}
    [\,p,q,r,s,t\,] \sim [\,p-2, q+1, r+1, s-2, t+1\,].
\end{equation}
In fact, this relation,
\begin{equation}
    R_0,R_1:\,
    \begin{pmatrix}
        I_2&&\\&r_3&\\&&r_3
    \end{pmatrix},
    \begin{pmatrix}
        I_2&&\\&r_3&\\&&r_3
    \end{pmatrix}
    \sim
    \begin{pmatrix}
        I_2&&\\&r_3&\\&&r_3
    \end{pmatrix},
    \begin{pmatrix}
        r_3&&\\&r_{-3}&\\&&I_2
    \end{pmatrix},
\end{equation}
is realized by the following SO(6) gauge transformation,
\begin{equation}
\begin{aligned}
    R'_0 &= \Omega(\omega z) R_0\, \Omega^\top (z), \\
    R'_1 &= \Omega(\omega z+1) R_1\, \Omega^\top (z),
\end{aligned}
\end{equation}
with
\begin{equation}
    \Omega(z)=\exp{\left[ \frac{2\pi}{3} J(a,b) \right]},\quad
    J(a,b)=
    \left(
    \begin{array}{cc|cc|cc}
0&0&-b&a&b&-a\\
0&0&-a&-b&-a&-b\\
\hline
b&a&0&0&-b&-a\\
-b&b&0&0&-a&b\\
\hline
-b&a&b&a&0&0\\
a&b&a&-b&0&0
\end{array}\right),
\end{equation}
where $z=a+bi$ $(a,b\in\mathbb{R})$.
$J(a,b)^\top=-J(a,b)$ is easily checked.
We emphasize that the equivalent relation (\ref{Z3_equiv}) is a consequence of the general classification based on the TCLs, meaning that no other relations exist.

Finally, let us count the number of the ECs.
The total patterns of the canonical forms (\ref{Z3_StandForm}), $\alpha_N$, are counted as
\begin{equation}
    \alpha_N= {}_{[\frac{N}{2}]+4} C_4 =
    \begin{cases}
        {}_{\frac{N}{2}+4} C_4 &\quad\text{for $N$= even,} \\
        {}_{\frac{N-1}{2}+4} C_4 &\quad\text{for $N$= odd,}
    \end{cases}
\end{equation}
where $[A]$ represents the greatest integer less than or equal to $A\in\mathbb{R}$.
From the equivalent relation (\ref{Z3_equiv}), it follows that some ECs contain several canonical forms and the overcounts are $\alpha_{N-6}$ $(N\geq6)$.
Thus, the total number of the ECs, $S_N$, is calculated as
\begin{equation}
    S_N=\alpha_N - \alpha_{N-6}=
    \begin{cases}
        \frac{1}{16}(N+2)(N^2+4N+8) 
        &\quad\text{for $N$= even,} \\
        \frac{1}{16}(N+1)(N^2+2N+5)
        &\quad\text{for $N$= odd.}
    \end{cases}
\end{equation}
Note that $\alpha_{N-6}=0$ for $N=1,\cdots,5$.
Table \ref{tab_Z3} summarizes the classification results of the BCs for $G=SO(N)$ in the $T^2/Z_3$ model.

\begin{table}[t]
    \centering
    \scalebox{0.9}{
    \renewcommand{\arraystretch}{1.8}
    \setlength{\tabcolsep}{10pt}
    \begin{tabular}{lc} 
    \hline \hline \\[-18pt]
    The canonical forms: &
    $\begin{alignedat}{50}
    R_0&= I_p 
    &&\oplus(I_2 \oplus\cdots)
    &&\oplus(r_3 \oplus\cdots)
    &&\oplus(r_3 \oplus\cdots)
    &&\oplus(r_3 \oplus\cdots) \\
    R_1&= I_p 
    &&\oplus \underbrace{(r_3 \oplus\cdots)}_{q}
    &&\oplus \underbrace{(I_2 \oplus\cdots)}_{r}
    &&\oplus \underbrace{(r_3 \oplus\cdots)}_{s}
    &&\oplus \underbrace{(r_{-3} \oplus\cdots)}_{t}
\end{alignedat}$ \\[18pt]
    The equivalent relations: &
    $[\,p,q,r,s,t\,] \sim [\,p-2, q+1, r+1, s-2, t+1\,]$ \\[9pt]
    The total number of the ECs: &
    $S_N=
    \begin{cases}
        \frac{1}{16}(N+2)(N^2+4N+8) 
        &\quad\text{for $N$= even} \\
        \frac{1}{16}(N+1)(N^2+2N+5)
        &\quad\text{for $N$= odd}
    \end{cases}$ \\[12pt]
    \hline \hline
    \end{tabular}}
    \caption{The classification results of the BCs in the $T^2/Z_3$ model with $G=SO(N)$.}
    \label{tab_Z3}
\end{table}

\subsection{\texorpdfstring{$T^2/Z_4$}{T2/Z4} orbifold}
Let us derive the canonical forms of the representation matrices $(R_0,R_1)$ for $G=SO(N)$ in the $T^2/Z_4$ model using the re-orthogonalization method.
$R_i \in O(N)$ $(i=0,1)$ satisfy the following consistency conditions (see Table \ref{tab_consis}):
\begin{equation}
    R_0^4=I_N,\quad R_1^4=I_N,\quad R_1 R_0 R_1 R_0=I_N.
\end{equation}
$(R_0, R_1)$ is simplified to the following form through unitary and gauge transformations \cite{Kawamura2023}:
\begin{equation} \label{Z4_diag}
\begin{alignedat}{20}
    R_0
    &=+I_p &&\oplus +I_q &&\oplus -I_r &&\oplus -I_s
    &&\oplus +i I_{t_1} &&\oplus +i I_{u_1} &&\oplus -i I_{u_2} &&\oplus -i I_{t_2}
    &&\oplus i^k\, (\sigma_3 \oplus\cdots),\\
    R_1
    &=+I_p &&\oplus -I_q &&\oplus +I_r &&\oplus -I_s
    &&\oplus +i I_{t_1} &&\oplus -i I_{u_1} &&\oplus +i I_{u_2} &&\oplus -i I_{t_2}
    &&\oplus i^{k-1} \underbrace{(\sigma_2 \oplus\cdots)}_{v},
\end{alignedat}
\end{equation}
where $k=0,1$ and $\sigma_i$ $(i=1,2,3)$ are the Pauli matrices.
$v$ denotes the number of each block.
An orthogonal matrix has the same numbers of conjugate complex eigenvalues, which leads to $t_1=t_2$ and $u_1=u_2$.
Since the complex conjugate eigenvalue pairs occupy the same positions in both $R_0$ and $R_1$, the re-orthogonalization method can be applied to Eq.(\ref{Z4_diag}).
Using the unitary matrix $V$ in Eq.(\ref{Re_V}), each block of $(R_0,R_1)$ in Eq.(\ref{Z4_diag}) are re-orthogonalized to
\begin{equation}
    V^\dag \begin{pmatrix} i &0\\0& -i \end{pmatrix} 
    V = r_4 \in O(2), \quad
    E^\top V^\dag \begin{pmatrix} 0 & -i \\ i& 0 \end{pmatrix} 
    V E= \sigma_3 \in O(2),
\end{equation}
with
\begin{equation}
    E= \frac{1}{\sqrt{2}}
    \begin{pmatrix}
        1 &1\\-1&1
    \end{pmatrix} \in O(2),
\end{equation}
where $r_m$ is defined in Eq.(\ref{r_m}).
As a result, we obtain the following two types of the canonical forms of $(R_0,R_1)$ for $G=SO(N)$:
\begin{align} 
\text{(i)}&:\quad \label{Z4_StandForm_k=0}
\begin{alignedat}{20}
    R_0
    &=+I_p &&\oplus +I_q &&\oplus -I_r &&\oplus -I_s
    &&\oplus (r_4 \oplus\cdots)
    &&\oplus (r_4 \oplus\cdots)
    &&\oplus (\sigma_3 \oplus\cdots),\\
    R_1
    &=+I_p &&\oplus +I_q &&\oplus -I_r &&\oplus -I_s
    &&\oplus \underbrace{(r_4 \oplus\cdots)}_{t}
    &&\oplus \underbrace{(r_{-4} \oplus\cdots)}_{u}
    &&\oplus \underbrace{(r_4 \oplus\cdots)}_{v},
\end{alignedat} \\
\text{(ii)}&:\quad \label{Z4_StandForm_k=1}
\begin{alignedat}{20}
    R_0
    &=+I_p &&\oplus +I_q &&\oplus -I_e &&\oplus -I_s
    &&\oplus (r_4 \oplus\cdots)
    &&\oplus (r_4 \oplus\cdots)
    &&\oplus (r_4 \oplus\cdots),\\
    R_1
    &=+I_p &&\oplus +I_q &&\oplus -I_e &&\oplus -I_s
    &&\oplus \underbrace{(r_4 \oplus\cdots)}_{t}
    &&\oplus \underbrace{(r_{-4} \oplus\cdots)}_{u}
    &&\oplus \underbrace{(\sigma_3 \oplus\cdots)}_{v},
\end{alignedat}
\end{align}
where $t(=t_i)$ and $u(=u_i)$ denote the numbers of each block.
Types (i) and (ii) correspond to the cases of $k=0$ and $k=1$ in Eq.(\ref{Z4_diag}).
Note that both types have the same form except for the last $v$ non-commutative blocks.

Examine the equivalent relations between the canonical forms through the TCLs.
First, we focus on the following gauge invariant quantities from the traces of $R_0^2$ and $R_1^2$:
\begin{align}
    \mathrm{tr} R_0^2 &= p+q+r+s-2t-2u\pm 2v, \\
    \mathrm{tr} R_1^2 &= p+q+r+s-2t-2u\mp 2v,
\end{align}
where the upper and lower signs denote Types (i) and (ii).
It is found that $v$ is invariant, that is, the non-commutative blocks in Eq.(\ref{Z4_StandForm_k=0}) and Eq.(\ref{Z4_StandForm_k=1}) are unchanged through the BCs-connecting gauge transformations.
Thus, we now focus on the commutative parts.
Since $R_0$ and $R_1$ commute with each other, only the following four traces are independent (see Table \ref{tab_tr}):
\begin{align}
    \mathrm{tr} R_0 &= p+q-r-s, \\
    \mathrm{tr} R_1 &= p-q+r-s, \\
    \mathrm{tr} R_0^2 &= p+q+r+s-2t-2u, \\
    \mathrm{tr} (R_1 R_0) &= p-q-r+s-2t+2u,
\end{align}
where $v$ is omitted.
Given that $p+q+r+s+2t+2u$ is invariant by definition, it follows that $p+q$, $p+r$, $p-s$, $p-t$ and $p+u$ are invariant.
Therefore, it is concluded that the possible transformations are restricted to the following:
\begin{equation} \label{Z4_equiv}
    [\,p,q,r,s \,|\,t,u \,|\, v\,] \sim [\,p-1,q+1,r+1,s-1 \,|\,t-1,u+1 \,|\, v\,].
\end{equation}
In fact, this relation,
\begin{equation}
    R_0,R_1:\,
    \begin{pmatrix}
        \sigma_3 &\\& r_4
    \end{pmatrix},
    \begin{pmatrix}
        \sigma_3 &\\& r_4
    \end{pmatrix}
    \sim
    \begin{pmatrix}
        \sigma_3 &\\& r_4
    \end{pmatrix},
    \begin{pmatrix}
        -\sigma_3 &\\& r_{-4}
    \end{pmatrix},
\end{equation}
is realized by the following SO(4) gauge transformation,
\begin{equation}
\begin{aligned}
    R'_0 &= \Omega(iz) R_0\, \Omega^\top (z), \\
    R'_1 &= \Omega(iz+1) R_1\, \Omega^\top (z),
\end{aligned}
\end{equation}
with
\begin{equation}
    \Omega(z)=\exp{\left[ \frac{\pi}{\sqrt{2}} J(a,b) \right]},\quad
    J(a,b)=
    \left(
    \begin{array}{cc|cc}
0&0&a-b&a+b\\
0&0&a-b&-a-b\\
\hline
-a+b&-a+b&0&0\\
-a-b&a+b&0&0\\
\end{array}\right),
\end{equation}
where $z=a+bi$ $(a,b\in\mathbb{R})$.
$J(a,b)^\top=-J(a,b)$ is easily checked.
We emphasize that the equivalent relation (\ref{Z4_equiv}) is a consequence of the general classification based on the TCLs, meaning that no other relations exist.

Finally, let us count the number of the ECs.
In the case of $v=0$, the total patterns of the canonical forms, $\alpha^{v=0}_{N}$, are counted as
\begin{equation}
    \alpha^{v=0}_{N}= \sum_{l=0}^{[N/2]} {}_{N-2l+3} C_3\cdot {}_{l+1}C_1.
\end{equation}
From the equivalent relation (\ref{Z4_equiv}), it follows that some ECs contain several canonical forms and the overcounts are $\alpha^{v=0}_{N-4}$ $(N\geq4)$.
The number of the ECs, $s_N$, is counted as $s_N=\alpha^{v=0}_{N}-\alpha^{v=0}_{N-4}$.
Note that $\alpha^{v=0}_{N-4}=0$ for $N=1,2,3$.
For $v\geq1$, there are two types of the non-commutative blocks: Type (i) and Type (ii).
Therefore, the total number of the ECs, $S_N$, is calculated as
\begin{align}
    S_N &= s_N +2\cdot \sum_{v=1}^{[N/2]} s_{N-2v} \\
    &=
    \begin{cases}
        \frac{1}{120}(N+2)(N^4+8N^3+34N^2+72N+60) 
        &\quad\text{for $N$= even,} \\
        \frac{1}{120}(N+1)(N+2)(N+3)(N^2+4N+15)
        &\quad\text{for $N$= odd.}
    \end{cases}
\end{align}
Table \ref{tab_Z4} summarizes the classification results of the BCs for $G=SO(N)$ in the $T^2/Z_4$ model.

\begin{table}[t]
    \centering
    \scalebox{0.9}{
    \renewcommand{\arraystretch}{1.8}
    \setlength{\tabcolsep}{10pt}
    \begin{tabular}{ll} 
    \hline \hline
    \multicolumn{2}{l}{The canonical forms (Types (i)-(ii)):} \\[6pt]
    \multicolumn{2}{c}{
    $\begin{aligned}
    \text{(i)}&:\quad
    \begin{alignedat}{20}
    R_0
    &=+I_p &&\oplus +I_q &&\oplus -I_r &&\oplus -I_s
    &&\oplus (r_4 \oplus\cdots)
    &&\oplus (r_4 \oplus\cdots)
    &&\oplus (\sigma_3 \oplus\cdots)\\
    R_1
    &=+I_p &&\oplus +I_q &&\oplus -I_r &&\oplus -I_s
    &&\oplus \underbrace{(r_4 \oplus\cdots)}_{t}
    &&\oplus \underbrace{(r_{-4} \oplus\cdots)}_{u}
    &&\oplus \underbrace{(r_4 \oplus\cdots)}_{v}
    \end{alignedat} \\[6pt]
    \text{(ii)}&:\quad
    \begin{alignedat}{20}
    R_0
    &=+I_p &&\oplus +I_q &&\oplus -I_r &&\oplus -I_s
    &&\oplus (r_4 \oplus\cdots)
    &&\oplus (r_4 \oplus\cdots)
    &&\oplus (r_4 \oplus\cdots)\\
    R_1
    &=+I_p &&\oplus +I_q &&\oplus -I_r &&\oplus -I_s
    &&\oplus \underbrace{(r_4 \oplus\cdots)}_{t}
    &&\oplus \underbrace{(r_{-4} \oplus\cdots)}_{u}
    &&\oplus \underbrace{(\sigma_3 \oplus\cdots)}_{v}
    \end{alignedat}
    \end{aligned}$}\\[18pt]
    The equivalent relations: &
    \scalebox{0.95}{
    $[\,p,q,r,s \,|\,t,u \,|\, v\,] \sim [\,p-1,q+1,r+1,s-1 \,|\,t-1,u+1 \,|\, v\,]$} \\
    \multicolumn{2}{l}{The total number of the ECs:} \\[6pt]
    \multicolumn{2}{c}{$S_N=
    \begin{cases}
        \frac{1}{120}(N+2)(N^4+8N^3+34N^2+72N+60) 
        &\quad\text{for $N$= even} \\
        \frac{1}{120}(N+1)(N+2)(N+3)(N^2+4N+15)
        &\quad\text{for $N$= odd}
    \end{cases}$} \\[12pt]
    \hline \hline
    \end{tabular}}
    \caption{The classification results of the BCs in the $T^2/Z_4$ model with $G=SO(N)$.}
    \label{tab_Z4}
\end{table}

\subsection{\texorpdfstring{$T^2/Z_6$}{T2/Z6} orbifold}
Let us derive the canonical forms of the representation matrices $(R_0,R_1)$ for $G=SO(N)$ in the $T^2/Z_6$ model using the re-orthogonalization method.
$R_i \in O(N)$ $(i=0,1)$ satisfy the following consistency conditions (see Table \ref{tab_consis}):
\begin{equation}
    R_0^6=I_N,\quad R_1^3=I_N,\quad R_1 R_0 R_1 R_0=I_N.
\end{equation}
$(R_0, R_1)$ is simplified to the following form by unitary and gauge transformations: \cite{Kawamura2023, 10.1093/ptep/ptae082}
\begin{equation} \label{Z6_diag}
\begin{aligned}
    R_0 &=\mathrm{diag}\,
    (\overbrace{1,\cdots}^{p},\overbrace{-1,\cdots}^{q} \,|\, 
    \overbrace{\eta,\cdots}^{r_1}, 
    \overbrace{\bar{\omega},\cdots}^{s_1} \,|\, 
    \overbrace{\omega, \cdots}^{s_2}, 
    \overbrace{\bar{\eta},\cdots}^{r_2},) \\[3pt]
    &\qquad\qquad\qquad\qquad\oplus \eta^k (A_0\oplus\cdots) \oplus \eta^{k+1} (A_0\oplus\cdots) \oplus \eta^l (B_0\oplus\cdots),
    \\[6pt]
    R_1 &=\mathrm{diag}\,
    (1,\cdots,\,\,1,\cdots \,|\, \omega,\cdots, \omega,\cdots \,|\, \bar{\omega},\cdots,  \bar{\omega},\cdots,) \\[3pt]
    &\qquad\qquad\qquad\qquad\oplus 
    \eta^{2k} \underbrace{(A_1\oplus\cdots)}_{t_1} \oplus
    \eta^{2(k+1)} \underbrace{(A_1\oplus\cdots)}_{t_2} \oplus 
    \eta^{2l} \underbrace{(B_1\oplus\cdots)}_{u},
\end{aligned}
\end{equation}
with
\begin{align}
    A_0&=
    \begin{pmatrix} -1&0\\0&1 \end{pmatrix}, \quad
    A_1= 
    \begin{pmatrix} 
    -\frac{1}{2}&\frac{\sqrt{3}}{2}i\\
    \frac{\sqrt{3}}{2}i&-\frac{1}{2} 
    \end{pmatrix}, \\[3pt]
    B_0&=
    \begin{pmatrix} \omega&&\\&\bar{\omega}&\\&&1 \end{pmatrix}, \quad
    B_1=
    \begin{pmatrix} -\frac{1}{3}\bar{\omega}& \frac{2}{3}\omega& \frac{2}{3} \\
    \frac{2}{3}\bar{\omega} & -\frac{1}{3}\omega & \frac{2}{3}\\
    \frac{2}{3}\bar{\omega} & \frac{2}{3}\omega & -\frac{1}{3} 
    \end{pmatrix},
\end{align}
where $\eta=e^{2\pi i/6}$, $k=0,1,2$ and $l=0,1$.%
\footnote{The off-diagonal components of $A_1$ are written as $\pm\sqrt{3}i/2$ in Ref.\!\cite{Kawamura2023}, but the double signs are equivalent by the unitary transformation using $\sigma_3$, so that it is sufficient to treat one of them.}
Let $p,q,r_i,s_i$ $(i=1,2)$ be the numbers of each eigenvalue and $t_i,u$ $(i=1,2)$ be the numbers of each block.
An orthogonal matrix has the same numbers of conjugate complex eigenvalues, which leads to
\begin{align}
    \text{for}\,\,k&=0:\quad r_1=r_2,\,s_1=s_2,\,t_2=0, \\
    \text{for}\,\,k&=1:\quad r_1=r_2,\,s_1=s_2,\,t_1=t_2, \\
    \text{for}\,\,k&=2:\quad r_1=r_2,\,s_1=s_2,\,t_1=0.
\end{align}
Since the complex conjugate eigenvalue pairs occupy the same positions in both $R_0$ and $R_1$, the re-orthogonalization method can be applied to Eq.(\ref{Z6_diag}).
Using the unitary matrix $V$ in Eq.(\ref{Re_V}), the diagonal parts in Eq.(\ref{Z6_diag}) are re-orthogonalized to
\begin{equation}
    V^\dag \begin{pmatrix} \eta &0\\0& \bar{\eta} \end{pmatrix} 
    V = r_6 \in O(2), \quad
    V^\dag \begin{pmatrix} \bar{\omega} &0\\0& \omega \end{pmatrix} 
    V = r_{-3} =-r_6 \in O(2),
\end{equation}
where $r_m$ is defined in Eq.(\ref{r_m}).
The pairs of the $2\times 2$ off-diagonal blocks in Eq.(\ref{Z6_diag}) are re-orthogonalized to
\begin{align}
    &\text{for}\,\,k=0:\quad \label{k=0}
    V^\dag V V_1^\dag (A_0,A_1) V_1 V^\dag V
    =(\sigma_3,r_3) \in O(2), \\
    &\text{for}\,\,k=1:\quad \label{k=2}
    V_4^\dag V_3^\dag (\eta A_0 \oplus \eta^2 A_0 \,,\, 
    \eta^2 A_1 \oplus \eta^4 A_1) V_3 V_4
    = (r_{-3}\otimes \sigma_3 \,,\, r_3 \otimes r_3)
    \in O(4), \\
    &\text{for}\,\,k=2:\quad 
    V^\dag V V_2^\dag (-A_0, A_1) V_2 V^\dag V
    =(\sigma_3,r_3) \in O(2), 
\end{align}
where
\begin{equation}
    V_1=
    \begin{pmatrix} 0&1 \\ -i&0 \end{pmatrix},\quad
    V_2=
    \begin{pmatrix} 1&0 \\ 0&i \end{pmatrix},\quad
    V_3
    \begin{pmatrix} I_2&0\\0&\sigma_1 \end{pmatrix}, \quad
    V_4= V\otimes I_2.
\end{equation}
The pairs of the $3\times 3$ off-diagonal blocks in (\ref{Z6_diag}) are re-orthogonalized to
\begin{equation}
    V_5^\dag E_l^\dag (\eta^l B_0, \eta^{2l} B_1) E_l V_5 =
    \left( \pm
    \begin{pmatrix}
        1& \\ & -r_6
    \end{pmatrix},
    \begin{pmatrix}
        -\frac{1}{3} & -\frac{2}{3} & -\frac{\sqrt{6}}{3}\\
        \frac{2\sqrt{2}}{3} & -\frac{1}{6} & -\frac{3}{6} \\
        0 & -\frac{3}{2} & \frac{1}{2}
    \end{pmatrix} \right)
    \equiv (\pm C_0,C_1) \in O(3),
\end{equation}
with
\begin{equation}
    E_0=
    \begin{pmatrix} &&1\\ &1&\\1&&  \end{pmatrix},\quad
    E_1=
    \begin{pmatrix} 1&& \\ &&1 \\ &1&  \end{pmatrix},\quad
    V_5=
    \begin{pmatrix} 1& \\ &V  \end{pmatrix},
\end{equation}
where the upper and lower signs correspond to $l=0$ and $l=1$.
As a result, the following four types of the canonical forms of $(R_0,R_1)$ for $G=SO(N)$ are obtained:
\begin{align}
\text{(i) :}&\quad \label{Z6_StandForm_1}
\begin{alignedat}{20}
    R_0 &= +I_p &&\oplus -I_q
    &&\oplus ( r_6 \oplus \cdots )
    &&\oplus ( -r_6 \oplus \cdots )
    &&\oplus (\sigma_3 \oplus \cdots)
    &&\oplus (+C_0 \oplus \cdots), \\
    R_1 &= +I_p &&\oplus +I_q
    &&\oplus \underbrace{( r_3 \oplus \cdots )}_{r}
    &&\oplus \underbrace{( r_3 \oplus \cdots )}_{s}
    &&\oplus \underbrace{(r_3 \oplus \cdots)}_{t}
    &&\oplus \underbrace{(C_1 \oplus \cdots)}_{u},
\end{alignedat} \\[6pt]
\text{(ii) :}&\quad \label{Z6_StandForm_2}
\begin{alignedat}{20}
    R_0 &= +I_p &&\oplus -I_q
    &&\oplus ( r_6 \oplus \cdots )
    &&\oplus ( -r_6 \oplus \cdots )
    &&\oplus (\sigma_3 \oplus \cdots)
    &&\oplus (-C_0 \oplus \cdots), \\
    R_1 &= +I_p &&\oplus +I_q
    &&\oplus \underbrace{( r_3 \oplus \cdots )}_{r}
    &&\oplus \underbrace{( r_3 \oplus \cdots )}_{s}
    &&\oplus \underbrace{(r_3 \oplus \cdots)}_{t}
    &&\oplus \underbrace{(C_1 \oplus \cdots)}_{u},
\end{alignedat} \\[6pt]
\text{(iii) :}&\quad \label{Z6_StandForm_3}
\begin{alignedat}{20}
    R_0 &= +I_p &&\oplus -I_q
    &&\oplus ( r_6 \oplus \cdots )
    &&\oplus ( -r_6 \oplus \cdots )
    &&\oplus (-r_6\otimes \sigma_3 \oplus \cdots)
    &&\oplus (+C_0 \oplus \cdots), \\
    R_1 &= +I_p &&\oplus +I_q
    &&\oplus \underbrace{( r_3 \oplus \cdots )}_{r}
    &&\oplus \underbrace{( r_3 \oplus \cdots )}_{s}
    &&\oplus \underbrace{(r_3\otimes r_3 \oplus \cdots)}_{t}
    &&\oplus \underbrace{(C_1 \oplus \cdots)}_{u},
\end{alignedat} \\[6pt]
\text{(iv) :}&\quad \label{Z6_StandForm_4}
\begin{alignedat}{20}
    R_0 &= +I_p &&\oplus -I_q
    &&\oplus ( r_6 \oplus \cdots )
    &&\oplus ( -r_6 \oplus \cdots )
    &&\oplus (-r_6\otimes \sigma_3 \oplus \cdots)
    &&\oplus (-C_0 \oplus \cdots), \\
    R_1 &= +I_p &&\oplus +I_q
    &&\oplus \underbrace{( r_3 \oplus \cdots )}_{r}
    &&\oplus \underbrace{( r_3 \oplus \cdots )}_{s}
    &&\oplus \underbrace{(r_3\otimes r_3 \oplus \cdots)}_{t}
    &&\oplus \underbrace{(C_1 \oplus \cdots)}_{u},
\end{alignedat}
\end{align}
where $r(=r_i)$, $s(=s_i)$ and $t(=t_i)$ denote the numbers of each block.
Note that Types (i)-(iv) have the same form except for the last $t$ and $u$ non-commutative blocks.

Examine the equivalent relations between the canonical forms through the TCLs.
First, we focus on the following gauge invariant quantities from the traces of $R_0^2$ and $R_1$:
\begin{align}
    \mathrm{tr} R_0^2 &= p+q-r-s \pm 2t, \\
    \mathrm{tr} R_1 &= p+q-r-s \mp t,
\end{align}
where the upper and lower signs denote Types (i)(ii) and (iii)(iv).
It is found that $t$ is invariant.
Next, the traces of $R_0^3$ and $(R_1 R_0)$ lead to
\begin{align}
    \mathrm{tr} R_0^3 &= p-q-2r+2s \pm 3u, \\
    \mathrm{tr} (R_1 R_0) &= p-q-2r+2s \mp u,
\end{align}
where the upper and lower signs denote Types (i)(iii) and (ii)(iv).
Here $t$ is omitted.
They mean that $u$ is also invariant, that is, the non-commutative blocks in Eqs.(\ref{Z6_StandForm_1})-(\ref{Z6_StandForm_4}) are unchanged through the BCs-connecting gauge transformations.%
\footnote{It also indicates that the non-commutative blocks cannot be reduced into smaller blocks.}
Therefore, we now focus on the commutative parts.
Since $R_0$ and $R_1$ commute with each other, only the following three traces are independent (see Table \ref{tab_tr}):
\begin{align}
    \mathrm{tr} R_0 &= p-q-r-s, \\
    \mathrm{tr} R_1 &= p+q-r-s, \\
    \mathrm{tr} (R_1 R_0) &= p-q-2r+2s,
\end{align}
where $t$ and $u$ are omitted.
Given that $p+q+2r+2s$ is invariant by definition, it follows that all $p$, $q$, $r$ and $s$ are invariant.
Therefore, it is concluded that the canonical forms (\ref{Z6_StandForm_1})-(\ref{Z6_StandForm_4}) are unchanged under the BCs-connecting gauge transformations.
In other words, Each EC contains only one canonical form in the $T^2/Z_6$ model.

Finally, let us count the number of the ECs.
the patterns of the canonical forms are counted as
\begin{align}
    \text{for}\,\,t=0,\, u=0 &: 
    \alpha^{(0,0)}_{N}= \sum_{l=0}^{[N/2]} {}_{N-2l+1} C_1\cdot {}_{l+1}C_1, \\
    \text{for}\,\,t=0,\, u\geq 1 &: 
    \alpha^{(0,u\geq 1)}_{N} = 2 \sum_{u=1}^{[N/3]} \alpha^{(0,0)}_{N-3u}\,, \\
    \text{for}\,\,t\geq 1,\, u=0 &: 
    \alpha^{(t\geq 1,0)}_{N} = 
    \sum_{t=1}^{[N/2]} \alpha^{(0,0)}_{N-2t} +
    \sum_{t=1}^{[N/4]} \alpha^{(0,0)}_{N-4t}\,, \\
    \text{for}\,\,t\geq 1,\, u\geq 1 &: 
    \alpha^{(t\geq 1, u\geq 1)}_{N} =
    2 \sum_{u=1}^{[N/3]} \alpha^{(t\geq 1,0)}_{N-3u},
\end{align}
where $l=r+s$.
The total number of the ECs, $S_N$, is written as
\begin{equation}
    S_N = \alpha^{(0,0)}_{N} + \alpha^{(0,u\geq 1)}_{N}
    + \alpha^{(t\geq 1,0)}_{N} + \alpha^{(t\geq 1, u\geq 1)}_{N}.
\end{equation}
As an example, $S_{10}=433$ for $G=SO(10)$.
Table \ref{tab_Z6} summarizes the classification results of the BCs in the $T^2/Z_6$ model.

\begin{table}[t]
    \centering
    \scalebox{0.9}{
    \renewcommand{\arraystretch}{1.8}
    \setlength{\tabcolsep}{10pt}
    \begin{tabular}{ll} 
    \hline \hline
    \multicolumn{2}{l}{The canonical forms (Types (i)-(iv)):} \\[6pt]
    \multicolumn{2}{c}
        {$\begin{aligned}
        \text{(i) :}&\quad 
        \begin{alignedat}{20}
        R_0 &= +I_p &&\oplus -I_q
        &&\oplus ( r_6 \oplus \cdots )
        &&\oplus ( -r_6 \oplus \cdots )
        &&\oplus (\sigma_3 \oplus \cdots)
        &&\oplus (+C_0 \oplus \cdots) \\
        R_1 &= +I_p &&\oplus +I_q
        &&\oplus \underbrace{( r_3 \oplus \cdots )}_{r}
        &&\oplus \underbrace{( r_3 \oplus \cdots )}_{s}
        &&\oplus \underbrace{(r_3 \oplus \cdots)}_{t}
        &&\oplus \underbrace{(C_1 \oplus \cdots)}_{u}
        \end{alignedat} \\[6pt]
        \text{(ii) :}&\quad 
        \begin{alignedat}{20}
        R_0 &= +I_p &&\oplus -I_q
        &&\oplus ( r_6 \oplus \cdots )
        &&\oplus ( -r_6 \oplus \cdots )
        &&\oplus (\sigma_3 \oplus \cdots)
        &&\oplus (-C_0 \oplus \cdots) \\
        R_1 &= +I_p &&\oplus +I_q
        &&\oplus \underbrace{( r_3 \oplus \cdots )}_{r}
        &&\oplus \underbrace{( r_3 \oplus \cdots )}_{s}
        &&\oplus \underbrace{(r_3 \oplus \cdots)}_{t}
        &&\oplus \underbrace{(C_1 \oplus \cdots)}_{u}
        \end{alignedat} \\[6pt]
        \text{(iii) :}&\quad 
        \begin{alignedat}{20}
        R_0 &= +I_p &&\oplus -I_q
        &&\oplus ( r_6 \oplus \cdots )
        &&\oplus ( -r_6 \oplus \cdots )
        &&\oplus (-r_6\otimes \sigma_3 \oplus \cdots)
        &&\oplus (+C_0 \oplus \cdots) \\
        R_1 &= +I_p &&\oplus +I_q
        &&\oplus \underbrace{( r_3 \oplus \cdots )}_{r}
        &&\oplus \underbrace{( r_3 \oplus \cdots )}_{s}
        &&\oplus \underbrace{(r_3\otimes r_3 \oplus \cdots)}_{t}
        &&\oplus \underbrace{(C_1 \oplus \cdots)}_{u}
        \end{alignedat} \\[6pt]
        \text{(iv) :}&\quad 
        \begin{alignedat}{20}
        R_0 &= +I_p &&\oplus -I_q
        &&\oplus ( r_6 \oplus \cdots )
        &&\oplus ( -r_6 \oplus \cdots )
        &&\oplus (-r_6\otimes \sigma_3 \oplus \cdots)
        &&\oplus (-C_0 \oplus \cdots) \\
        R_1 &= +I_p &&\oplus +I_q
        &&\oplus \underbrace{( r_3 \oplus \cdots )}_{r}
        &&\oplus \underbrace{( r_3 \oplus \cdots )}_{s}
        &&\oplus \underbrace{(r_3\otimes r_3 \oplus \cdots)}_{t}
        &&\oplus \underbrace{(C_1 \oplus \cdots)}_{u}
        \end{alignedat} 
        \end{aligned}$} \\[9pt]
    The equivalent relations:
    & $[p,q,r,s \,|\,t, u]$ is gauge invariant. \\
    The total number of the ECs: &
    $S_N = \alpha^{(0,0)}_{N} + \alpha^{(u\geq 1,0)}_{N}
    + \alpha^{(0,t\geq 1)}_{N} + \alpha^{(u\geq 1, t\geq 1)}_{N}$ \\[6pt]
    \hline \hline
    \end{tabular}}
    \caption{The classification results of the BCs in the $T^2/Z_6$ model with $G=SO(N)$.}
    \label{tab_Z6}
\end{table}

\section{Conclusion and discussion} \label{sec_concl}
We have classified the equivalence classes (ECs) of the $T^2/Z_m$ $(m=2,3,4,6)$ orbifold boundary conditions (BCs) for the gauge group $G=SO(N)$.
The canonical forms of the representation matrices have been derived through the ``re-orthogonalization method."
Next we have examined all the possible equivalent relations between the canonical forms by using the gauge invariant quantities obtained from the trace conservation laws (TCLs).
Finally, the numbers of the ECs have been calculated.
The classification results for each orbifold model are summarized in Tables \ref{tab_Z2}-\ref{tab_Z6}.

As shown in the results, the canonical forms for $G=SO(N)$ are quite different from the ones for $G=SU(N)$.
In the case of $SU(N)$, the representation matrices $R_i$ can be simultaneously diagonalized in the $T^2/Z_2$ and $T^2/Z_3$ models, but in the $T^2/Z_4$ and $T^2/Z_6$ models, $R_i$ cannot always be simultaneously diagonalized and possess the off-diagonal blocks.
This arises from the $Z_2$ or $Z_3$ sub-symmetry of the $T^2/Z_4$ and $T^2/Z_6$ models.
On the other hand, in the case of $SO(N)$, the canonical forms remain diagonal in the $T^2/Z_2$ model, but the forms in the $T^2/Z_3$, $T^2/Z_4$ and $T^2/Z_6$ models are generally block-diagonal. 
This is due to that the orthogonal matrices $R_i$ have the complex eigenvalues in these models and cannot be diagonalized while keeping their orthogonality.
In such models, it is known that the rank of the representation matrices are reduced without the continuous Wilson line phases, leading to phenomenologically interesting symmetry breaking \cite{Kawamura2023}.
It would be fascinating to search the $T^2/Z_3$ models with the off-diagonal blocks, which do not exist in the case of $SU(N)$.

Even in the case of $SO(N)$, there are non-trivial equivalent relations between the canonical forms, which strongly suggests that the Hosotani mechanism works.
Especially, the $T^2/Z_3$ and $T^2/Z_4$ models show characteristic behavior, where the canonical forms transition to other canonical forms without ever being diagonalized.
It is interesting to study how the AB phase causes the spontaneous symmetry breaking dynamically in these models.

The TCLs have a wide range of applications because they are valid regardless of the gauge groups and the shapes of orbifolds.
In previous and current works, we have achieved the general classification of the ECs for $G=SU(N)$ and $SO(N)$.
Recently, the gauge-Higgs unification model with $G=Sp(6)$ has been studied phenomenologically \cite{Maru:2024ver}.
The TCLs are applied to such models with various compact Lie groups.
In addition, the TCLs also play a central role in the higher-dimensional orbifold models and the magnetized orbifold models \cite{10.1093/ptep/ptw058, PhysRevD.96.096011, IMAI2023116189, Kikuchi2023, PhysRevD.108.036005, Kojima2023, 10.1093/ptep/ptae070, Kobayashi2024NI}.

As seen in Tables \ref{tab_Z2}-\ref{tab_Z6}, the ECs still contain arbitrariness.
There have been several attempts to address the arbitrariness problem of the BCs\cite{YAMAMOTO201445}, but a convincing mechanism or principle for determining BCs is still unclear.
As an approach to this problem, we propose classifying the ECs in the blow-up manifolds of the orbifolds\cite{kobayashi2019, kobayashi2020, PhysRevD.107.075032}.
These are smooth manifolds with the fixed points removed, which make it possible to analyze the critical phenomena around the fixed points in a well-defined manner.
Studying the ECs in the blow-up procedure would reveal the relationships between the ECs.
We hope to report on them in the future.

\section*{Acknowledgement}
The authors would like to thank Yoshiharu Kawamura, Kenta Kojin and Gaia Higa for the insightful discussions.
The authors thank the Yukawa Institute for Theoretical Physics at Kyoto University, where this work was initiated during the conference YITP-W-24-09 on "Progress in Particle Physics 2024".
This work was supported by JST SPRING, Grant Number JPMJSP2132.

\bibliographystyle{unsrt} 
\bibliography{main} 

\end{document}